\documentclass{article}

\PassOptionsToPackage{numbers, compress}{natbib}

\usepackage[final]{nips_2018}

\usepackage[utf8]{inputenc} 
\usepackage[T1]{fontenc}    
\usepackage{hyperref}       
\usepackage{url}            
\usepackage{booktabs}       
\usepackage{amsfonts}       
\usepackage{nicefrac}       
\usepackage{microtype}      
\usepackage{amsmath}
\usepackage[pdftex]{graphicx} 

\title{Modeling Rape Reporting Delays Using Spatial, Temporal and Social Features}

\author{
  Konstantin Klemmer\textsuperscript{*}, Daniel B. Neill\textsuperscript{\$} \& Stephen A. Jarvis\textsuperscript{*} \\
  \textsuperscript{*} Department of Computer Science, University of Warwick \& The Alan Turing Institute \\
  \textsuperscript{\$} Department of Computer Science \& Center for Urban Science and Progress,
  New York University \\
  \texttt{\{k.klemmer, s.a.jarvis\}@warwick.ac.uk, daniel.neill@nyu.edu}
}

\begin{document}

\maketitle

\begin{abstract}
We present a novel approach to estimate the delay observed between the occurrence and reporting of rape crimes. We explore spatial, temporal and social effects in sparse \textit{aggregated} (area-level) and high-dimensional \textit{disaggregated} (event-level) data for New York and Los Angeles. Focusing on inference, we apply Gradient Boosting and Random Forests to assess predictor importance, as well as Gaussian Processes to model spatial disparities in reporting times. Our results highlight differences and similarities between the two cities. We identify at-risk populations and communities which may be targeted with focused policies and interventions to support rape victims, apprehend perpetrators, and prevent future crimes.
\end{abstract}

\section{Introduction}

With the Harvey Weinstein scandal and the \texttt{\#metoo} hashtag campaign, 2017 brought the attention of the general public once again to the issue of sexual violence. While the last decades have shown progress on destigmatization, victims of sexual violence still suffer from substantial social repercussions. This manifests itself in extremely low reporting numbers for sexual crime, the most severe of which is rape. 
In 2012, a survey conducted by the parenting forum \textit{mumsnet} found that 10\% of women had been raped and 35\% had been sexually assaulted \cite{Mumsnet2012}. Compounding the problem is only a small proportion of rape crimes are reported to police: the Rape, Abuse \& Incest National Network (RAINN) estimates that only around 30\% of all sexual assault is reported \cite{RAINN2014}, and our analysis of New York City crime data (described below) reveals that approximately 30\% of rapes that were reported to police had a lag of 30 days or more between occurrence and reporting time. Given the substantial problem of under-reporting, it is of crucial importance to understand the underlying process of reporting sexual violence in general, and rape in particular. This issue is reiterated by a recent study that showed that the longer a victim hesitates to report, the more unlikely it becomes for the crime to be reported at all \cite{Bicanic2015}. Additionally, reporting delays complicate investigation efforts, making it more difficult to convict perpetrators. With this study, we seek to tackle the problem of delayed rape reporting from a machine learning (ML) perspective. In recent years, ML has emerged as a powerful toolkit for criminology research \cite{Brennan2013}, for instance helping to predict domestic violence \cite{Berk2016}. We apply Random Forests and Gradient Boosting to assess the importance of multiple socioeconomic and temporal factors in predicting reporting time. We observe substantial spatial dependence, thus applying Gaussian Processes to model the correlation structure and to examine whether there is residual spatial dependence, after controlling for neighborhood-level demographic and socioeconomic features.

This paper contributes to current research by providing a new dataset of predictors of rape reporting delays compiled from open data sources, exposing spatial variation in rape reporting and gathering novel insights into the importance of spatio-temporal and socioeconomic factors. Our research has the potential to motivate policies and support community policing practices. Subsequently, we introduce our dataset, test for spatial and temporal effects, present the results for our predictive models and finally suggest potential future research directions.

\section{Data}

We access historical crime records for New York City (NYC) and Los Angeles (LA), from their open-data portals. To ensure a sufficient sample size for regional aggregation, we extract rape crimes from 2012 to 2017. To ensure representativeness of our predictors, we select only rape which occurred in the domestic or professional space, i.e. where we can assume that victim and perpetrator were acquainted. This amounts to a total of 7,242 reported rapes for NYC and 4,649 for LA. Note that while we experimented with shorter time periods as well (due to the static nature of our predictors), model performance did not vary substantially. For each crime $i$ we compute the reporting delay in days $\Delta t_{i} = t^{occ}_{i} - t^{rep}_{i}$ as the difference between the day the crime occurred and the day it was reported to authorities. We then create binary values indicating whether a crime was reported within one day or within one month respectively:  
\begin{equation}
    d^{day}_{i}=
    \begin{cases}
      1, & \text{if}\ \Delta t_{i}  \leq 1 \\
      0, & \text{otherwise}
    \end{cases}
  \ \text{and} \
d^{month}_{i}=
\begin{cases}
    1, & \text{if}\ \Delta t_{i}  \leq 30 \\
      0, & \text{otherwise}
    \end{cases}
 \end{equation}
The LA crimes come with point location data for each rape case, normalized to block level. We can hence use these points without further processing for our \textit{disaggregated model}. The NYC data, however, comes with police precincts as the most granular geography. For each of the 77 NYC precincts $j$, we compute the proportion of crimes reported within the two selected intervals $p^{day}_{j} = \frac{1}{m_{j}} \sum_{i \in j} d^{day}_{i}$ and $p^{month}_{j} = \frac{1}{m_{j}} \sum_{i \in j} d^{month}_{i}$, where $m_{j}$ represents the number of observed rapes in each area. We also create an aggregated dataset for LA, aggregating on the police department's ''basic car district'' level (169 basic car districts in total).  Each of these three datasets (aggregated NYC, aggregated LA, disaggregated LA) were enriched with the following features: demographic (age, sex, race), housing (household size, occupants per room, single households, rent as percentage of income), education (high-school degrees) and economic (income, unemployment, insurance). For the disaggregated dataset, we collect statistics from the census tract within which the crime was committed. For the aggregated data, we apply population-weighted data aggregation from census tracts to police precincts (NYC) or basic car districts (LA). Lastly, the LA data provides information on the victim's age, gender and ethnicity which we use in the disaggregated model. 

\begin{figure}
  \centering
  \includegraphics[,scale=0.5]{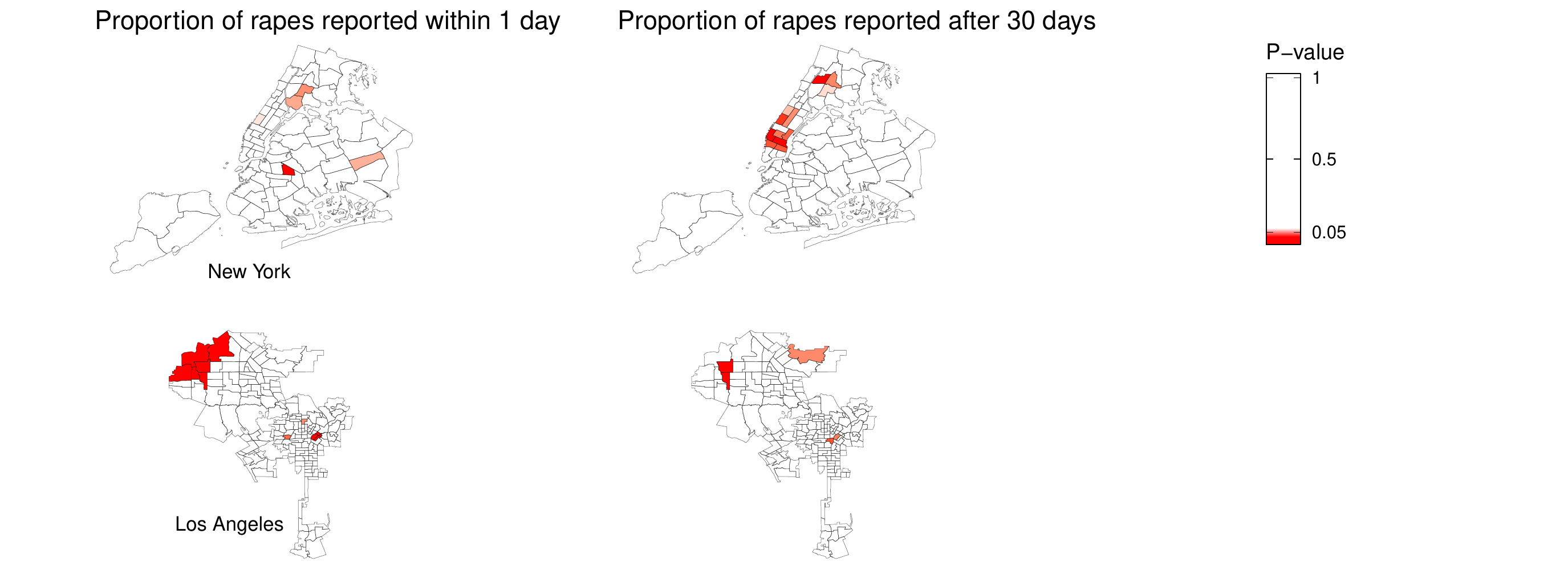}
  \caption{Local spatial autocorrelation (Moran's I) p-values of reporting delay proportions $p_{j}^{day}$ and $p_{j}^{month}$ in New York and Los Angeles.}
\end{figure}


\section{Results}

\subsection{Spatial and Temporal Effects}

We start our analysis by examining geographical and temporal patterns in rape reporting delays. Our hypothesis is that reporting delays are in fact not \textit{i.i.d.}, but rather exhibit spatial and temporal dependencies \cite{Melo2016}. We measure local spatial autocorrelation for the proportions $p_{j}$ of crimes reported within the intervals $\{day,month\}$ using the \textit{Local Moran's I} statistic \cite{Anselin1995}, defined as
\begin{equation}
    I_{j}= (n-1) \frac{p_{j}-\bar{p}}{\sum_{k=1,k \neq j}^{m} w_{j,k}(p_{k} - \bar{p})^{2}} \sum_{k=1,k \neq j}^{m} w_{j,k}(p_{k} - \bar{p})
\end{equation}
where $\bar{p}$ represents the mean of $p_{j}$ and $w_{j,k}$ represents the spatial weight between the features $j$ and $k$. We assume simple weights for neighbouring and non-neighbouring areas of each district $j$, where $w_{j,k}=1$ if $k$ is a neighbour of $j$ and $w_{j,k}=0$ otherwise.

The \textit{Local Moran's I} statistic corresponds to a (pseudo) p-value indicating the significance of the spatial correlation. Both NYC and LA exhibit hot spots of spatial autocorrelation, confirming our initial hypothesis (see Figure 1). There are (at least) two possible explanations for this: \textit{(1)} an underlying social process disguised as a spatial process (e.g., reporting delays are driven by socio-economic factors that vary over space), or \textit{(2)} a true spatial process (e.g., a contagious spread of rape reporting behavior over space). We address this question in the next section.

\begin{figure}
  \centering
  \includegraphics[,scale=0.5]{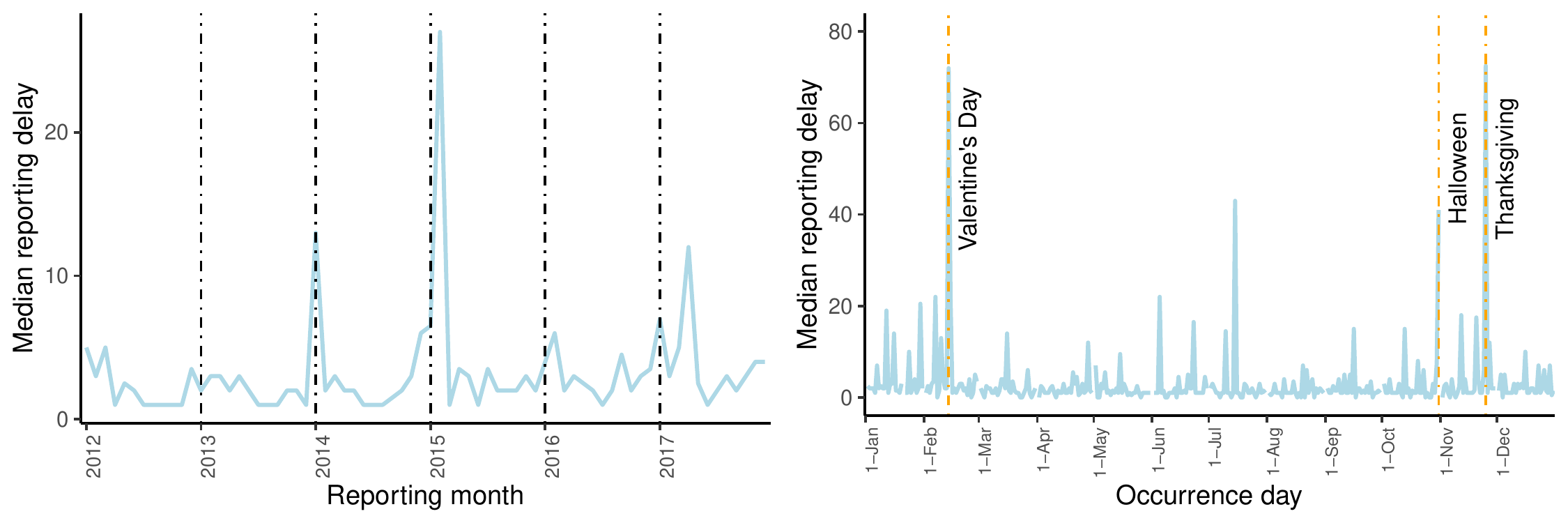}
  \caption{Median rape reporting delays for New York over all observed months (left) and all days of a year (right).}
\end{figure}

First, however, we assess the temporal dimension. Again, our hypothesis is that temporal features affect rape reporting delays. In particular, we examine whether rapes that occurred on weekends, holidays or during different seasons exhibit varying reporting delays. Figure 2 shows median reporting delays by respective reporting month and occurrence day. Several of the observed spikes and irregularities can be explained with common policing practices. For instance, for rapes where the victim cannot remember the exact occurrence date (due to trauma or long reporting delay), the first or middle day of a month is recorded as occurrence day. This practice is consistent for NYC and LA. However, we do find interesting patterns from the graphical analysis: Holidays (particularly those highlighted in Figure 2) exhibit higher median reporting delays. Note that this pattern of longer delays does not seem to correspond to more overall rapes during these days. Beyond holidays, rapes reported at the beginning and end of each year appear to be reported later. We hence decide to include both spatial and temporal features in our predictive model for reporting delays.  

\subsection{Predictive Modeling}

\textbf{Aggregated model:} We model the aggregated data in a two-step process. First, we estimate our output $p_j^{\{day,month\}}$ as a function of social features, $p \sim f(x) + \epsilon$ where $x$ represents a matrix of social features and $\epsilon$ gives the model residuals. We then seek to predict the residual error term $\epsilon \sim f(g)$ using spatial context, where $g$ represents the centroid coordinates of each area $j$ to isolate additional spatial effects that are not explained by the variation in observed demographic and socioeconomic features. We apply this to aggregate data from NYC and LA, excluding temporal features which are measured at disaggregated case level only. Step 1 is carried out using Random Forests and Gradient Boosting (choosing the best predictor by cross-validation each time), Step 2 uses Gaussian Processes with radial basis function (RBF) kernels to learn from spatial context \cite{Buhmann2003}. We apply Leave-One-Out Cross-Validation (LOO-CV) to validate the predictions. Our results are shown in Table 1. For NYC, we can explain around 20\% of the total variation in the data using social features. Spatial context explains another 10\% of the remaining variation. In LA, the explanatory power of social features drops slightly while spatial context does not further improve the predictions. Looking at variable importance from the random forest model, we find that economic predictors (e.g. median income, unemployment rate) are relatively unimportant, while housing characteristics (e.g. occupants per room, household size) appear highly relevant. African-American communities are associated with shorter reporting delays, while Hispanic communities are associated with longer reporting delays. These findings are consistent across both cities.

\begin{table}
\centering
\caption{Prediction results for the aggregated model with LOO-CV: MSE and $R^{2}$ values for the best performing models in Step 1 and Step 2}
\label{table1}
\begin{tabular}{lclcl|clll}
       & \multicolumn{4}{c}{New York}  & \multicolumn{4}{c}{Los Angeles} \\
       \hline
       & \multicolumn{2}{c}{$p_{j}^{day}$} & \multicolumn{2}{c|}{$p_{j}^{month}$} & 
       \multicolumn{2}{c}{$p_{j}^{day}$} & \multicolumn{2}{l}{$p_{j}^{month}$} \\
       \hline
       & MSE & $R^{2}$ & MSE & $R^{2}$ & MSE & $R^{2}$ & MSE & $R^{2}$  \\
       \hline
       \textbf{Step 1: $p \sim f(x)$} & 0.0049 & 0.243 & 0.005 & 0.233 & 0.0014 & 0.176 & 0.0049 & 0.097 \\
       \textbf{Step 2: $\epsilon \sim f(g)$} & 0.0044 & 0.09 & 0.0045 & 0.095 & 0.0013 & 0.035 & 0.0048  & 0.004                    
\end{tabular}
\end{table}

\textbf{Disaggregated model:} We model disaggregated data as a classification problem where we seek to predict each case's binary indicators $d_{i}^{\{day,month\}}$ as defined in section 2. These are modelled as a function $d \sim f(c,z,x,v)$ where $x$ represents social features for the census tract a rape occurred in, $z$ represents temporal features (holiday, weekend,...) specific to the occurrence date, $v$ represents victim-specific information and $c$ represent the coordinates of crime location (normalized on block level). We use Gaussian Process, Random Forest and Gradient Boosting classifiers to apply this model to the LA data. Due to the large sample size, we use $10$-fold cross validation instead of LOO-CV to verify the results, which are given in Table 2.

\begin{table}[!htbp]
\centering
\caption{Prediction results for the disaggregated model with $k=10$-fold CV: Log-Loss and AUC are given for the respectively best performing model.}
\label{table2}
\begin{tabular}{l|ll|ll|ll|ll|ll}
& \multicolumn{2}{c|}{$d \sim f(c)$} & \multicolumn{2}{c|}{$d \sim f(z)$} & \multicolumn{2}{c|}{$d \sim f(x)$} & \multicolumn{2}{c|}{$d \sim f(v)$} & \multicolumn{2}{c}{$d \sim f(c,z,x,v)$} \\
\hline
& LogL & AUC & LogL & AUC & LogL & AUC & LogL & AUC & LogL & AUC \\
\hline
$d_{i}^{day}$ & 0.682 & 0.521 & 0.678 & 0.531 & 0.681 & 0.531 & 0.65 & 0.626 & 0.656 & 0.625 \\
$d_{i}^{month}$ & 0.592 & 0.501 & 0.574 & 0.545 & 0.59 & 0.517 & 0.536 & 0.68 & 0.53 & 0.692             
\end{tabular}
\end{table}

The findings indicate that predictive power of spatial, temporal and social features is limited, while victim information adds substantial value to the model. We achieve our best prediction accuracy with an AUC of nearly 0.7 for $p_{i}^{month}$ from a model including all available features. Variable importance measures confirm findings from the aggregated model and also demonstrate that victim age is by far the best single predictor of reporting delay, which is in line with previous findings \cite{Smith2000}. Despite suspected patterns, temporal features do not appear to be of particular relevance for this analysis. 

\section{Conclusion}

In this paper, we present a machine learning approach to predict reporting delays of rape crime. We train event-level \textit{(disaggregated)} and area-level \textit{(aggregated)} models with spatial, temporal and socioeconomic predictors gathered for the cities of New York and Los Angeles. Our findings suggest that economic factors and temporal features have relatively low, and housing characteristics relatively high, explanatory power. We find substantial evidence for spatial patterns in rape reporting, which we account for using Gaussian Processes to model covariance structures. While these results may already serve authorities in tailoring policies and community-policing strategies, further efforts should be undertaken to deepen the understanding of rape reporting processes. Bayesian approaches coming with more generalizable inference might be particularly valuable. We explore the temporal dimension of rape reporting using static features only, which could be approved upon using a dynamic approach. This way, effects of events such as the \texttt{\#metoo} campaign could also be investigated. As only some of our socioeconomic predictors appear relevant, more features quantifying cultural influences (e.g., social stigma) should be tested. Lastly, working with open data sources limits the available information and is not fully representative of the data available to authorities. Working with detailed victim and perpetrator data might help to further isolate vulnerable subgroups who could be supported with targeted community-based interventions. Nevertheless the potential threat of victim reidentification from open data should also be addressed.

\subsubsection*{Acknowledgments}

The authors would like to thank Yu Chen, Matt Dwyer, Davey Ives, Rachel Lim and Sunglyoung Kim for their ideas, feedback and support. The authors gratefully acknowledge funding from the UK Engineering and Physical Sciences Research Council, the EPSRC Centre for Doctoral Training in Urban Science (EPSRC grant no. EP/L016400/1); The Alan Turing Institute (EPSRC grant no. EP/N510129/1).




\bibliographystyle{unsrt}
\bibliography{nips_2018}

\end{document}